\documentclass[prd,reprint,nofootinbib,notitlepage,aps,tightenlines,
preprintnumbers,amsmath,amssymb,showpacs,superscriptaddress]{revtex4-1}

\usepackage[hyperindex,breaklinks]{hyperref}
\usepackage{graphicx,epstopdf,amsmath,amsfonts,amssymb,color}

\def\be{\begin{equation}}
\def\ee{\end{equation}}
\def\ba{\begin{eqnarray}}
\def\ea{\end{eqnarray}}
\def\ge{\mathrel{\raise.3ex\hbox{$>$\kern-.75em\lower1ex\hbox{$\sim$}}}}
\def\la{\mathrel{\raise.3ex\hbox{$<$\kern-.75em\lower1ex\hbox{$\sim$}}}}

\def\simgt{\mathrel{\raise.3ex\hbox{$>$\kern-.75em\lower1ex\hbox{$\sim$}}}}
\def\simlt{\mathrel{\raise.3ex\hbox{$<$\kern-.75em\lower1ex\hbox{$\sim$}}}}

\newcommand{\bi}[1]{\bibitem{#1}}
\newcommand{\fr}[2]{\frac{#1}{#2}}

\newcommand{\nc}{\newcommand}

\nc{\gone}{\bar g_{\pi NN}^{(1)}}
\nc{\gzero}{\bar g_{\pi NN}^{(0)}}
\nc{\al}{\alpha}
\nc{\ga}{\gamma}
\nc{\de}{\delta}
\nc{\ep}{\epsilon}
\nc{\ze}{\zeta}
\nc{\et}{\eta}
\nc{\ka}{\kappa}
\nc{\rh}{\rho}
\nc{\si}{\sigma}
\nc{\ta}{\tau}
\nc{\up}{\upsilon}
\nc{\ph}{\phi}
\nc{\ch}{\chi}
\nc{\ps}{\psi}
\nc{\om}{\omega}
\nc{\Ga}{\Gamma}
\nc{\De}{\Delta}
\nc{\La}{\Lambda}
\nc{\Si}{\Sigma}
\nc{\Up}{\Upsilon}
\nc{\Ph}{\Phi}
\nc{\Ps}{\Psi}
\nc{\Om}{\Omega}
\nc{\ptl}{\partial}
\nc{\del}{\nabla}
\nc{\ov}{\overline}
\nc{\newcaption}[1]{\centerline{\parbox{15cm}{\caption{#1}}}}
\nc{\us}{U(1)$_S$}
\nc{\Rg}{$R_{\gamma\gamma}$}

\def\beq{\begin{equation}}
\def\eeq{\end{equation}}
\def\bmat{\begin{displaymath}}
\def\emat{\end{displaymath}}
\def\bear{\begin{eqnarray}}
\def\eear{\end{eqnarray}}
\def\ba{\begin{eqnarray}}
\def\ea{\end{eqnarray}}
\def\bery{\begin{array}}
\def\ery{\end{array}}
\def\bit{\begin{itemize}}
\def\eit{\end{itemize}}
\def\ben{\begin{enumerate}}
\def\een{\end{enumerate}}
\def\btab{\begin{tabular}}
\def\etab{\end{tabular}}
\def\btbl{\begin{table}}
\def\etbl{\end{table}}
\def\bfig{\begin{figure}[htb]}
\def\efig{\end{figure}}
\def\bpic{\begin{picture}}
\def\epic{\end{picture}}

\def\ga{\mathrel{\raise.3ex\hbox{$>$\kern-.75em\lower1ex\hbox{$\sim$}}}}
\def\la{\mathrel{\raise.3ex\hbox{$<$\kern-.75em\lower1ex\hbox{$\sim$}}}}
\def\gappeq{\mathrel{\rlap {\raise.5ex\hbox{$>$}}
{\lower.5ex\hbox{$\sim$}}}}
\def\lappeq{\mathrel{\rlap{\raise.5ex\hbox{$<$}}
{\lower.5ex\hbox{$\sim$}}}}

\def\gyr{{\rm \, G\kern-0.125em yr}}
\def\mev{{\rm \, Me\kern-0.125em V}}
\def\gev{{\rm \, Ge\kern-0.125em V}}
\def\tev{{\rm \, Te\kern-0.125em V}}

\def\lsim{\mathrel{\rlap{\lower4pt\hbox{\hskip1pt$\sim$}}
    \raise1pt\hbox{$<$}}}                % less than or approx. symbol
\def\gsim{\mathrel{\rlap{\lower4pt\hbox{\hskip1pt$\sim$}}
    \raise1pt\hbox{$>$}}}                % greater than or approx. symbol

\begin{document}
 
\title{CKM benchmarks for electron EDM experiments}

\author{Maxim Pospelov}
\affiliation{Department of Physics and Astronomy, University of Victoria, 
Victoria, BC V8P 5C2, Canada}
\affiliation{Perimeter Institute for Theoretical Physics, Waterloo, ON N2J 2W9, 
Canada}

\author{Adam Ritz}
\affiliation{Department of Physics and Astronomy, University of Victoria, 
Victoria, BC V8P 5C2, Canada}

\date{November 2013}

\begin{abstract}
All current experiments searching for an electron EDM $d_e$ are performed with 
atoms and diatomic molecules.  Motivated by significant recent progress in searches for an EDM-type signal in diatomic molecules with 
an uncompensated electron spin, we provide an estimate for the expected signal in the Standard Model due to the CKM phase. 
We find that the main contribution originates from the effective electron-nucleon operator $\bar e i\gamma_5 e \bar NN$, 
induced by a combination of weak and electromagnetic interactions at $O(G_F^2\alpha^2)$, 
and not by the CKM-induced electron EDM itself. When the resulting atomic $P,T$-odd mixing is interpreted as an {\em equivalent} electron EDM, 
this estimate leads to the benchmark $d_e^{\rm equiv}({\rm CKM}) \sim 10^{-38}e{\rm cm}$.

\end{abstract}
\maketitle

\section{Introduction}
\label{sec:intro}

Electric dipole moments (EDMs) of nucleons, atoms and molecules, have for many years provided some of our most sensitive probes for new sources of $T$-violation in nature, as required for baryogenesis. There has been significant experimental progress in the past few years \cite{YbF,Hgnew,n,newlimit}, most recently with the announcement of an impressive limit on $T$-odd effects in the polar molecule ThO, interpreted as a stringent constraint on the electron EDM, $|d_e| < 8.7 \times 10^{-29} e{\rm cm}$ \cite{newlimit}. The new physics reach of these experiments varies depending on the source of $T$ (or $CP$) violation, but can reach 100's of TeV \cite{PR}. 

The Standard Model (SM) itself has two sources of $CP$-violation: the Kobayashi-Maskawa phase, characterized by the reduced Jarlskog invariant ${\cal J}={\rm Im}(V_{tb}V^*_{td}V_{cd} V^*_{cb})\sim 3\times10^{-5}$ \cite{jarlskog}, and $\theta_{\rm QCD} = \theta_0 - {\rm Arg}(Y_u Y_d)$ which is constrained by the limit on the neutron EDM to be below $10^{-10}$ \cite{n}. Since the value of $\theta_{\rm QCD}$ is unknown, for the purposes of this paper we will treat it as a source of new physics, and focus our attention on the Kobayashi-Maskawa phase \cite{km} in the Cabbibbo-Kobayashi-Maskawa (CKM) mixing matrix, which is now well tested as the dominant source of $CP$-violation in the kaon and $B$-meson system. Given the continuing improvements in experimental sensitivity to EDMs, it is natural to ask about the size of the observable EDMs, $d_{\rm obs}$, induced by the CKM phase. In practice, calculations of these contributions are not available with high precision, and the estimates represent in effect the ultimate level of sensitivity of EDM searches to new physics. One can turn this statement around and ask, given the limited calculational precision available for $d_{\rm obs}({\cal J})$, what is the largest conceivable size of these CKM-induced contributions? Answering this question would assist us in defining a `line in the sand', corresponding to the level at which a nonzero EDM detection would unambiguously be due to new physics. However, the difficulty in quantifying the size of CKM-induced EDMs is apparent on noting that similar physical mechanisms, e.g. penguin diagrams in the up-quark sector, are at play in evaluating $CP$-odd observables in kaon physics. Specifically in the case of $\ep'/\ep$, the hadronic matrix elements are enhanced by factors of ${\cal O}(10)$ compared to naive expectations. 

CKM contributions to a number of observable EDMs have been discussed in the literature, as we will review below. However, the case of atoms and molecules where  the angular momentum is carried by an uncompensated electron spin
(leading to what is often referred to as a `paramagnetic EDM') 
has not been explored in detail. Paramagnetic EDMs are an important class of observables, allowing for relatively 
precise theoretical calculations of the dependence on underlying $CP$-odd sources of new physics, such as the electron EDM. Providing an estimate of the CKM contribution, and thus the effective threshold for EDM searches, is the main goal of this work and the result is summarized below.

First, we define the electron EDM operator $d_e$, and the semileptonic $CP$-odd operator $C_{SP}$,
\be
\label{LCP}
 {\cal L}_{CP} = -\frac{i}{2} d_e \bar{e} F\sigma \gamma_5 e - \frac{G_F}{\sqrt 2} C_{SP} \bar{N} N \bar{e} i \gamma_5 e + \cdots
\ee
$C_{SP}$ does not depend on the nuclear spin, gives a contribution to the atomic/molecular EDM even for spin-zero nuclei, and is coherently enhanced by the number of nucleons $A$. This singles  out $C_{SP}$ as likely the most important contribution to 
paramagentic EDMs among a multitude of other $CP$-odd four-fermion operators. In general $C_{SP}$ can have isospin dependence, 
which for this paper we will disregard, taking $C_{SP}$ to be an approximate isoscalar. 
As defined above, with the Fermi constant factored out, $C_{SP}$ is dimensionless.

One can write the shift of atomic/molecular energy levels under an applied external field ${\cal E}_{\rm ext}$ as 
\be
\label{DE}
\frac{\Delta E}{{\cal E}_{\rm ext}} = f_d (d_e + r C_{SP}) + \cdots
\ee
The coefficient $f_d$ reflects the relativistic violation of the Schiff theorem, and provides large enhancement factors \cite{Sandars} for heavy paramagnetic atoms, and particularly for polarizable paramagnetic molecules.\footnote{Molecular polarization is a nonlinear function of the applied electric field, and so $f_d$ itself is a nonlinear function of ${\cal E}_{\rm ext}$. 
}  The coefficient  $r$ in (\ref{DE}) has the dimensions of a dipole, $e$cm, 
and is determined by a ratio of the atomic matrix elements of the $C_{SP}$ and $d_e$ operators. 
Over the years, significant theoretical effort has gone into computing the $f_d$ and $r$ coeffecients for different molecular and 
atomic species; see {\it e.g.} \cite{lk,df,mkt,dfh,mb}.

 If only one species is used for an EDM measurement, the 
effects of $C_{SP}$ and $d_e$ cannot be separated (see {\it e.g.} \cite{jung,ermk} for recent discussions). Since the experimental sensitivity is usually reported as an inferred limit on the electron 
EDM, it is convenient to parametrize the effect of $C_{SP}$ as a contribution from an {\em equivalent} EDM:
\be
\label{EDMeq}
d_e^{\rm equiv} \equiv r C_{SP}.
\ee
Taking the three leading experimental limits on the electron EDM, we list the relevant $r$ coefficients \cite{lk,df,mkt,dfh,mb}, 
\begin{align}
r_{\rm Tl} &=  1.2\times 10^{-20} e{\rm cm},\nonumber\\
r_{\rm YbF} &= 0.88\times 10^{-20}e{\rm cm},\label{rcoeff}\\
r_{\rm ThO} &= 1.33\times 10^{-20}e{\rm cm}\nonumber.
\end{align} 
Notice that although the $f_d$ coefficients for these systems actually differ widely, the $r$ coefficients 
are approximately the same, reflecting the very similar dynamical nature of the $P,T$-odd perturbations 
to the electron Hamiltonian generated by both terms in (\ref{LCP}). This leaves 
only mild species-dependence in $d_e^{\rm equiv}$.

 In this paper, we find that in the Standard Model the CKM-induced $C_{SP}$ contribution dominates the direct contribution from $d_e$, and 
estimate it as 
\be
 C_{SP}({\cal J}) \sim 10^{-18},
 \label{result}
\ee
where ${\cal J}$ is the reduced Jarlskog invariant. Using the $r$ coefficients in (\ref{rcoeff}), we can translate this into a characteristic CKM background to searches for the electron EDM,
\be
 d_e^{\rm equiv} ({\cal J})\sim 10^{-38}e{\rm cm}.
\ee
This is roughly nine orders of magnitude below the best current sensitivity to $d_e$, from ThO \cite{newlimit}. 

The rest of this paper is organized as follows. In the next section, we briefly review the CKM contributions to fundamental fermions and other observable EDMs. In Section~3, we turn to the CKM contribution to paramagnetic EDMs, and obtain the result (\ref{result}). We finish with some concluding remarks in Section~4.

\section{Overview of EDMs from the CKM phase}

In this section, we briefly review existing computations of EDMs induced by the CKM phase. We will organize the discussion around a simple counting scheme, using the basic symmetries to estimate the largest viable contribution to different classes of EDMs. In particular, CKM contributions to flavor-diagonal observables necessarily vanish at first order in the weak interaction, due to the conjugated weak vertices. Nonzero contributions only start at second order $\propto G_F^2$, and are necessarily proportional to the reduced Jarlskog invariant 
\be
{\cal J}  = s_1^2s_2s_3c_1c_2c_3 \sin\delta \simeq 2.9\times10^{-5},
\ee
where $s_i$ and $c_i$ are the sines and cosines of the CKM angles in the Kobayashi-Maskawa basis and $\delta$ is the complex phase. 
The antisymmetric flavour structure of ${\cal J}$ also leads to additional loop-level suppression of the EDMs of quarks and leptons in perturbation theory \cite{shabalin,kp} 

\subsection{Fundamental fermion EDMs}

In addition to the general constraints above, the EDM operator for quarks and leptons breaks chiral symmetry, and thus the coefficient must  be at least linear in a chirality breaking parameter. In perturbation theory, this is generically the fermion mass $m_f$ itself. It turns out that the antisymmetric flavour struture of ${\cal J}$ actually ensures that all 2-loop contributions to $d_q$ vanish \cite{shabalin}, and the second-order weak exchanges need to be dressed with a further gluonic loop. Thus, the $d$-quark EDM for example arises only at 3-loop order \cite{khriplovich,czarnecki}, and takes the general form
\begin{align}
\label{dq}
 d_d^{\rm (est)} ({\cal J}) &\sim e {\cal J} \frac{\al_s\al_W^2}{(4\pi)^3} \frac{m_d}{m_W^2} \frac{m_c^2}{m_W^2}< 10^{-34} e{\rm cm}.
\end{align}
This estimates assigns $\alpha_i/(4\pi)$ per corresponding loop, and ignores 
additional numerical suppression or modest numerical enhancement by
 logarithms of quark  mass ratios. The factor of $m_c^2$ enters due to the flavour structure of ${\cal J}$. The corresponding contribution to $d_u$ is instead proportional to $m_um_s^2$, and somewhat further suppressed. The most precise calculation of $d_q({\cal J})$ can be found in Ref. \cite{czarnecki}. 

\begin{figure}
\includegraphics[viewport=160 450 440 720, clip=true, scale=0.4]{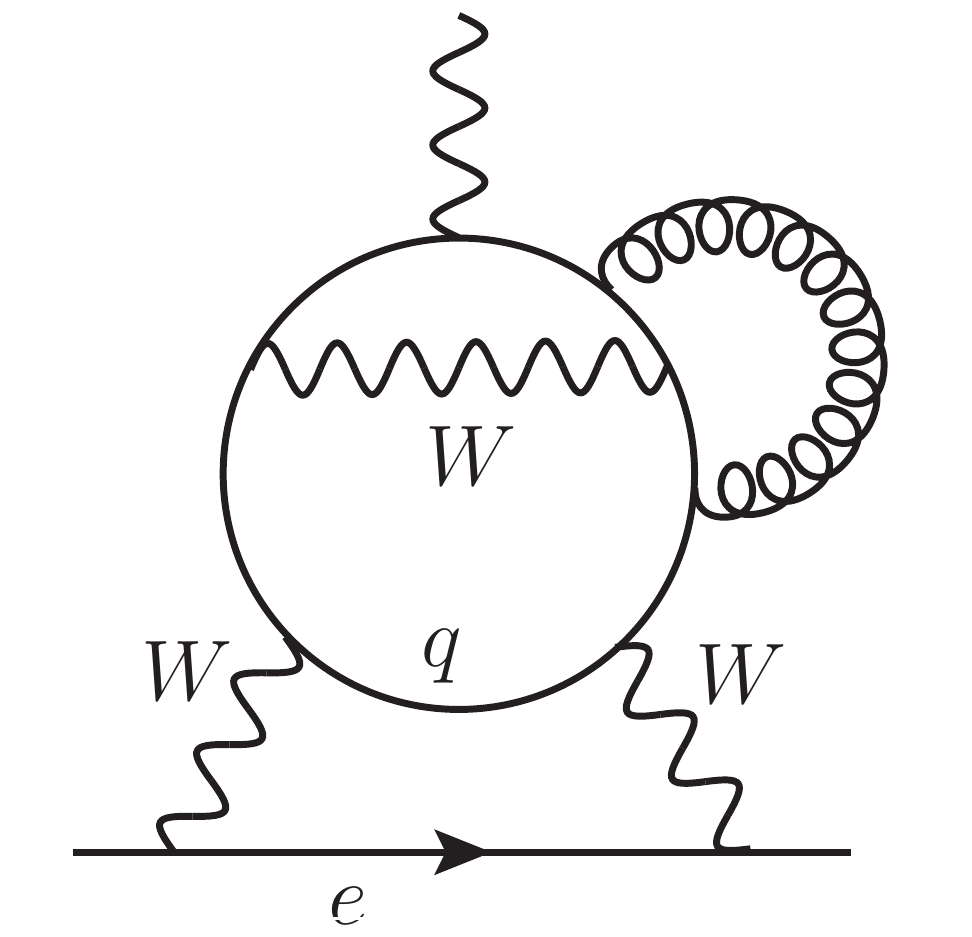}
\includegraphics[viewport=160 450 440 720, clip=true, scale=0.4]{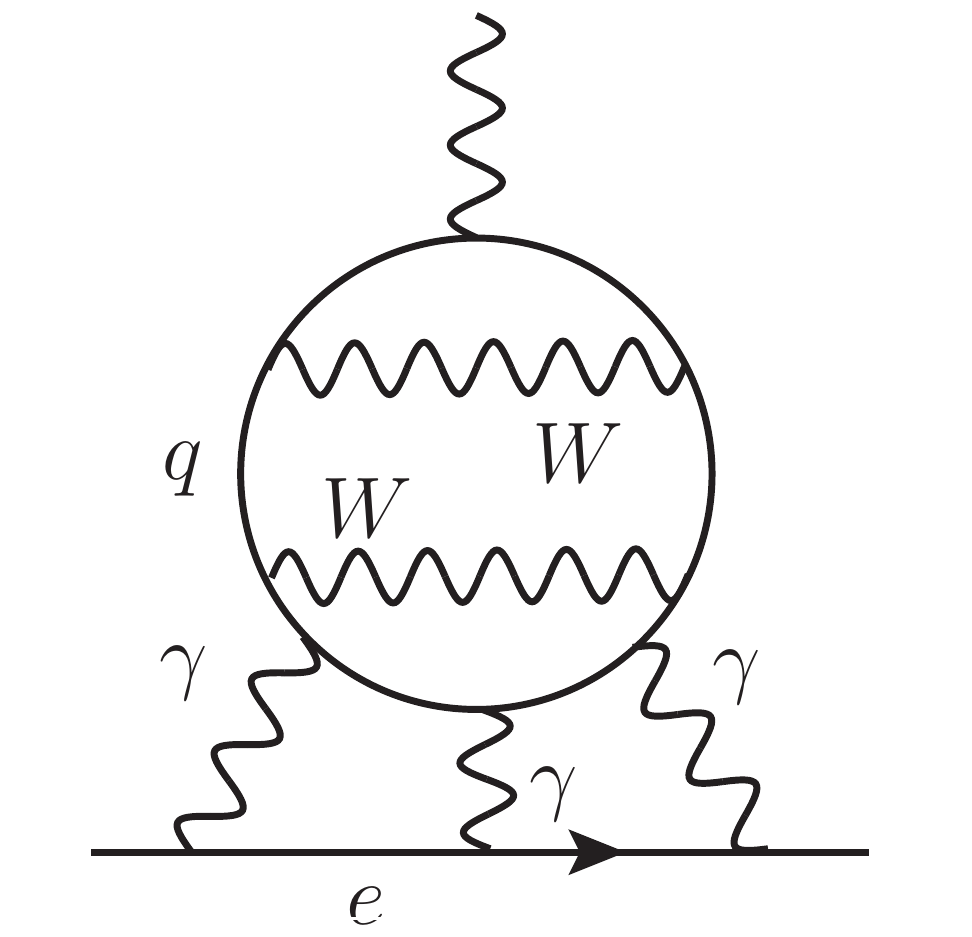}
\caption{Electron EDM $d_e$ induced by the CKM phase via a closed quark loop. The contributions shown are:
$O(\alpha_W^3\alpha_s)$ (left panel, Fig.~1a), and $O(\alpha^2\alpha^3)$ (right panel, Fig.~1b). } 
\label{fig1}
\end{figure}

EDMs of leptons are even further suppressed. A generic $d_e$ diagram involves a 
quark loop with a minimum of four $W$-boson vertices. Such a loop can be attached 
to the electron line either by two $W$-boson lines (Fig.~1a), at third order in the weak interaction,
or via three virtual photons (Fig.~1b), at even higher loop order.

Moreover, as mentioned before, the full 3-loop contribution to $d_e$ vanishes once 
again on account of the implicit antisymmetry of ${\cal J}$ \cite{kp}. An additional gluonic loop is required to generate $d_e$ at 4-loop order.  Thus, one can estimate the results for the 
two families of diagrams in Fig.~1, taking into account all the relevant coupling constants,
\begin{eqnarray}
\label{4loop}
d_e^{\rm Fig.1a} \sim e {\cal J} \frac{m_em_c^2m_s^2}{m_W^6}\frac{\alpha_W^3\alpha_s}{(4\pi)^4},\\
d_e^{\rm Fig.1b} \sim e{\cal J} \frac{m_em_c^2m_s^2}{m_W^4m_{\rm had}^2}\frac{\alpha^2_W\alpha^3}{(4\pi)^5},
\label{5loop}
\end{eqnarray} 
where in the second line $m_{\rm had}$ is a soft QCD mass scale ({\em e.g.} $m_\pi$) accounting for the 
fact that the loop may be saturated in the IR. To obtain an estimate, it suffices to take 
$m_{\rm had}\sim m_s$. Both contributions to $d_e$ are highly suppressed and give comparable values, 
\begin{align}
\label{EDMe}
d_e({\cal J}) &\sim O(10^{-44})~e{\rm cm}.
\end{align}
A very small number indeed!

To conclude this section, we discuss the origin of the 
quark mass suppression factors in Eqs.~(\ref{4loop}) and (\ref{5loop}) in more detail. These expressions contain an extra factor 
of $m_s^2$ compared to the $d$-quark EDM estimate, Eq.~(\ref{dq}). This 
factor originates from the closed quark loop of Fig.~1, where a complete antisymmetrization over $d,s,b$ 
quark masses is applied compared to the open quark line with the $d$ flavor as an in- and out-state, 
where only the $s$ and $b$ flavors internal to diagram are antisymmetrized. As a result,
the quark diagram can contain $m_s$, $m_b$ factors in a logarithm, $d_d \propto 
\log(m_b^2/m_s^2)$ and the power-like GIM suppression by $m_s^2$ is avoided. 
For the closed quark loop, complete antisymmetrization leads to $\log(m_b^2/m_s^2)+\log(m_d^2/m_b^2)+\log(m_s^2/m_d^2)=0$, and consequently the power-like 
GIM suppression by $m_s^2$ necessarily arises. Explicit calculations of the 
quark loop in the CKM model giving rise to the triple-gluon Weinberg operator \cite{PospelovW} and the magnetic quadrupole moment of the $W$-boson \cite{kp2} confirm the power-suppression by $m_s^2$. One can also argue that since Fig.~1a  is third order in the electroweak coupling, the $1/M_W^{6}$ factor is inevitable, as the $W$-bosons can be integrated out to give contact $\sim G_F$ interactions \cite{khriplovich}. Then a factor of $({\rm mass})^5$ is required in the numerator, and 
$m_em_s^2 m_c^2$ is the only combination of quark and electron masses that is consistent with all the symmetries of the problem. If, for instance, 
$m_s^2/M_W^2$ were to be absent, it would signal a quadratic divergence in the contact limit with loop momenta 
on the order of $M_W^2$. If that were possible, all down-type quarks could be considered massless and setting $m_s=m_d$ would nullify the answer. Retaining the finite $m_s^2/M_W^2$ correction returns us to the estimate (\ref{4loop}).

\subsection{Nucleon EDMs}

In practice, these primary fermion EDMs are not the dominant source of the CKM-induced EDMs of nucleons, and diamagnetic atoms. The largest CKM contributions generically arise through $CP$-odd multi-quark operators, containing (part of) the required flavour structure to produce ${\cal J}$. For 4-quark operators of this type, there is also the possibility of enhanced hadronic-scale contributions when these operators contribute to the interactions between nucleons and light pseudoscalar mesons. To get an idea of the size of possible enhancements, we can write down an expression for the nucleon EDM in a form which accounts, as above, for the irreducible requirements, and makes no further assumptions about small parameters,
\begin{align}
 d_N^{\rm (lim)}({\cal J}) &\sim e c_{n} {\cal J} G_F^2 m_{\rm had}^3 \nonumber\\
  &< 10^{-29} e{\rm cm} \times c_n\left(\frac{m_{\rm had}}{300\,{\rm MeV}}\right)^3.
\end{align}
In this limiting estimate, the chiral parameter $m_{\rm had}$ is taken to be characteristic of the quark condensate. In all practical estimates, the overall coefficient $c_n \ll 1$, but is not known with great precision.

\begin{figure}
\includegraphics[viewport=160 520 440 720, clip=true, scale=0.45]{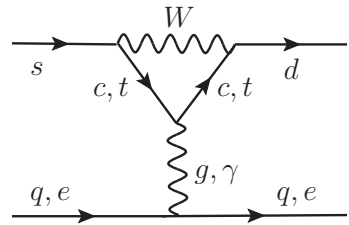}
\caption{Strong and electromagnetic penguin diagrams generating flavor-changing 
$CP$-violating four-quark and semi-leptonic operators. } 
\label{pen1}
\end{figure}

The antisymmetry of ${\cal J}$ requires that the operators obtained by combining weak currents at second order must contain at least two sea-quark flavours, e.g. $s$ and $c$. Integrating out the $c,t$ quarks at 1-loop via a strong penguin (see Fig.~\ref{pen1}) allows for the possibility of an enhanced phase, with the GIM cancelation factor being rather large, $\ln(m_t^2/m_c^2)$ (or rather $\ln(m_W^2/m_c^2)$ since $m_W^2<m_t^2$). This penguin-induced 4-quark operator was originally used by Khriplovich and Zhitnitsky to estimate $d_n({\cal J})$ via $CP$-odd $n\pi\Si$  and $nKp$ vertices entering a pion loop \cite{kz}. In these diagrams, the chiral scale $m_{\rm had}^3 \sim \langle \bar{q}q\rangle m_\pi^2/(f_\pi m_s)\sim (300\,{\rm MeV})^3$ is enhanced, while the coefficient $c_n$ roughly scales as $c_n \sim \al_s/(4\pi) \ln (m_K/m_\pi) \sim 10^{-2}$, leading to $d_n \sim 10^{-32}- 10^{-31} e$cm. A somewhat larger estimate, $d_n \propto10^{-30}  e {\rm cm}$, was obtained in Ref.~\cite{Gavela}. 

An alternative to generating 4-quark operators at 1-loop  is to integrate out charm at tree-level, generating a 6-quark operator with a coefficient $\propto 1/m_c^2$. Recently, evidence for enhanced $CP$-violation in the $D$ meson system led to further scrutiny of this contribution by Mannel and Uraltsev \cite{mu}. While avoiding the corresponding loop factor, there is suppression by $1/m_c^2$ and it remains difficult to obtain reliable estimates for the matrix element of this dimension-9 operator over the nucleon. Scaling estimates suggest $c_n \sim m_{\rm had}^2/m_c^2 \sim 10^{-2}$ again leading to $d_n \sim 10^{-31} e$cm \cite{mu}.

\subsection{Diamagnetic EDMs}

Atomic EDMs can be characterized in a similar way, but one needs to account for the Schiff theorem. We will focus on the case of paramagnetic systems below, but for diamagnetic atoms (such as Hg) Schiff screening suppresses the induced atomic EDM by a factor of roughly $10^3$, barring special cases with deformed nuclei. 
The leading contribution is determined by the Schiff moment $S$, and in addition to contributions from the individual nucleon EDMs, there is also the possibility of an enhanced contribution from the $CP$-odd nucleon potential. Indeed, the penguin diagram in Fig.~1 will contribute to $\bar{N}N \bar{N}i\gamma_5 N$ interactions and thus to the $CP$-odd nucleon potential, via e.g. kaon exchange. This mechanism was first studied in \cite{fks}, and reconsidered in \cite{dhm}, and may provide a contribution to $S({\cal J})$ which is somewhat larger than the nucleon EDMs. 
Focussing on $d_{\rm Hg}$, the dominant contribution takes the form \cite{fg},
\begin{align}
 d_{\rm Hg}({\cal J}) &\sim  -10^{-17} \left(\frac{S({\cal J})}{e{\rm fm}^3}\right) e {\rm cm} + \cdots \nonumber\\
  &\sim 10^{-25} \et_{np} ({\cal J}) e{\rm cm},
\end{align}
where $\et_{np}$ provides a dimensionless normalization of the 4-nucleon interactions; schematically  ${\cal L}_{\rm nuc} = \frac{1}{\sqrt 2} G_F \et_{np} \bar{N}N \bar{N}i\gamma_5 N$.
The precision of the nuclear calculation leading to the second line above is under scrutiny in the recent literature \cite{Engel} (see also \cite{ermk}), but will be sufficient for our discussion below.

An estimate for $\et_{np}({\cal J})$ can be obtained along the same lines as those above. Recalling that a factor $G_F$ forms part of the definition, we expect the leading contribution to emerge at first order in $G_F$, so the estimate takes the form,
\begin{align}
 \et_{np}^{({\rm lim})}({\cal J})&\sim c_{\rm Schiff} {\cal J}  G_F m_{\rm had}^2 \nonumber \\
  &\sim 10^{-11} \times c_{\rm Schiff}\left(\frac{m_{\rm had}}{300\,{\rm MeV}}\right)^2.
\end{align}
For $c_{\rm Schiff}\sim {\cal O}(1)$, this is roughly in line with the chiral constraints and matrix element estimates of Donoghue et al \cite{dhm}. This implies a CKM contribution to
$d_{\rm Hg}({\cal J}) < 10^{-35} e{\rm cm}$.

\section{CKM contribution to the electron-nucleon \boldmath{$CP$}-violating interaction}

We turn now to the main topic of this note, namely the CKM contribution to `paramagnetic' atoms and molecules containing an uncompensated electron spin. As discussed earlier, we are interested in the EDM-equivalent contribution from $C_{SP}$, Eq. (\ref{EDMeq}). 
The highly suppressed nature of $d_e({\cal J})$ reviewed above implies that the dominant CKM background arises from $C_{SP}$, which we will proceed to compute below. We first consider a simple scaling estimate along the same lines as those above. Accounting for the required chirality flips, and the factor of $e^4/16\pi^2 = \al^2$ 
required to connect electron and nucleon (or quark) lines with the minimal loop factor suppression, we have the following limiting value
\begin{align}
 C_{SP}({\cal J})^{\rm (lim)} &\sim c_C {\cal J} G_F \alpha^2m_e m_{\rm had} \nonumber\\
  &\sim 3\times 10^{-18}\times c_C \left(\frac{m_{\rm had}}{300\,{\rm MeV}}\right).
\label{est}
\end{align}
As we will discover shortly, a more elaborate estimate turns out to be about an order of magnitude lower than (\ref{est}) with $c_C \sim {\cal O}(1)$. 
However, it is already apparent that the CKM-induced $d_e^{\rm equiv}$ will not exceed $10^{-37}e$cm.

\begin{figure}
\includegraphics[viewport=160 500 450 720, clip=true, scale=0.4]{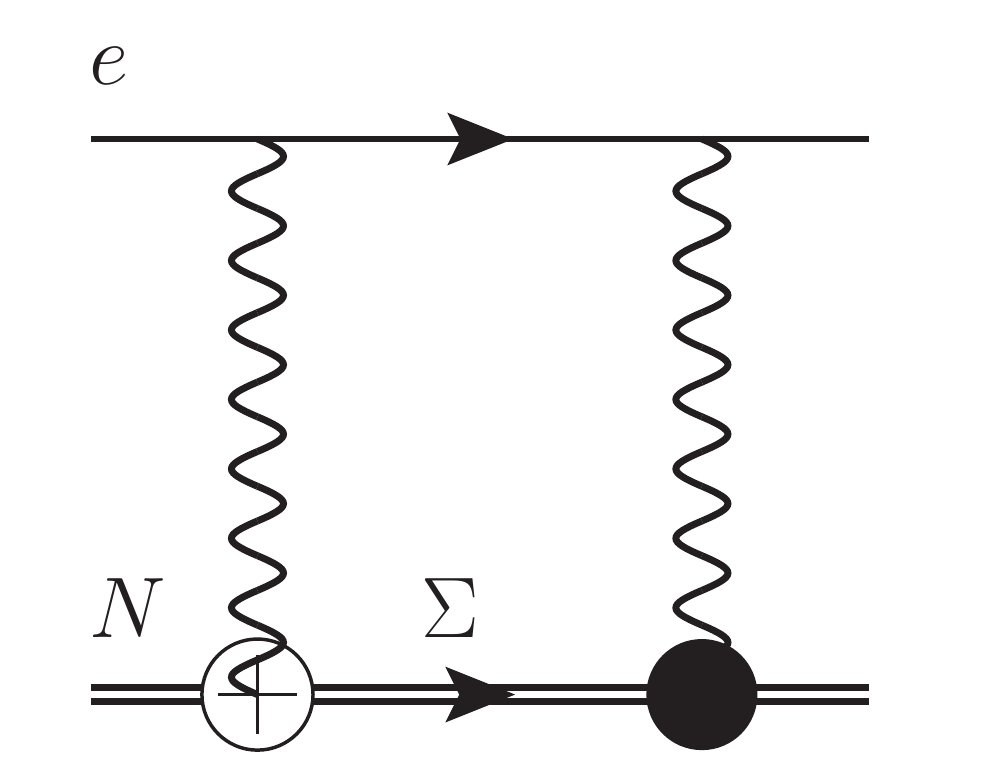}
\includegraphics[viewport=160 500 450 720, clip=true, scale=0.4]{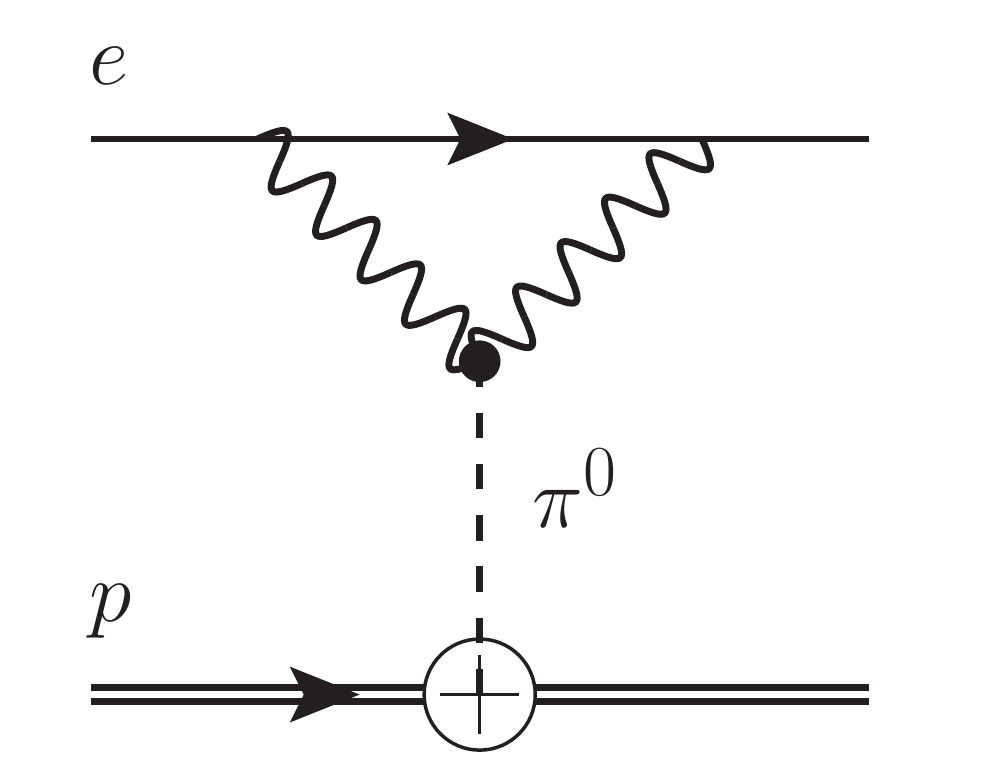}
\caption{Examples of the 2PE mechanism, leading to $\bar e i \gamma_5 e \bar NN$
interactions. Left panel (Fig.~3a): a combination of two weak transitions changing strangeness by $\pm1$.
The crossed and filled circles stand for the $CP$-odd and $CP$-even $\Sigma\-N\-\gamma$ vertices; the $CP$-odd vertex is induced by an EM penguin as in Fig.~2. 
Right panel (Fig.~3b): a diagram involving $\pi^0$ mediation. The crossed vertex in this case represents the  $CP$-odd $\pi^0NN$ coupling. } 
\label{figCS}
\end{figure}

\subsection{Effective electron-photon-nucleon \boldmath{$CP$}-odd operator}

The 2-photon exchanges (2PE) between the electron and the nucleon exhibited in Fig.~3 suggest
that it is natural to first identify the local $CP$-violating nucleon operator, obtained in the 
limit when the hadronic scales are considered to be 
larger than the virtuality $k$ of the photon loop. The 
leading dimension 2-photon operator is  $\bar NN F_{\mu\nu}\tilde F_{\mu\nu}$ 
(we ignore nuclear spin-dependent contributions such as $\bar Ni\gamma_5N F_{\mu\nu} F_{\mu\nu}$). The pion exchange diagram, shown in Fig.~3b, can be interpreted in precisely this form. We will instead focus on another important, and in some sense more useful, operator associated with Fig.~3a which arises as follows. Keeping in mind that the operators of interest will not couple to the nucleon spin, we start with the following higher-dimensional nucleon-photon operator associated with the lower part of Fig.~3a, $ \tilde{F}_{\mu\nu}(\partial_\alpha F_{\alpha\mu})\ptl_\nu( \bar{N}N)$, where $\tilde{F}_{\mu\nu} \equiv \ep_{\mu\nu\rh\si} F_{\rh\si}/2$. The factor of $\ptl_\al F_{\al\mu}$ is characteristic of the electromagnetic penguin, and can be traded immediately for the electron electromagnetic current, $\bar e \gamma_\mu e$. This singles out the following dimension-9 operator coupling the nucleon to the electron current,
\be
{\cal L}_{\rm eff} = eC_9 {\cal O}_9,~~~ {\cal O}_9 = \tilde F_{\mu\nu} (\bar e \gamma_\nu e) \partial_\mu (\bar NN).
\label{O9}
\ee
The operator ${\cal O}_9$, and thus the Wilson coefficient $C_9$, is $T$- and $P$-odd, and $C$-even, as required to contribute to $C_{SP}$.
Note that another possible nucleon-spin-independent operator $\tilde F_{\mu\nu} (\bar NN)\partial_\mu (\bar e \gamma_\nu e)$ is directly reducible to $O_9$ upon integration 
by parts and use of the identity $\partial_\mu \tilde F_{\mu\nu} \equiv 0$. 

\begin{figure}
\includegraphics[viewport=160 520 450 720, clip=true, scale=0.4]{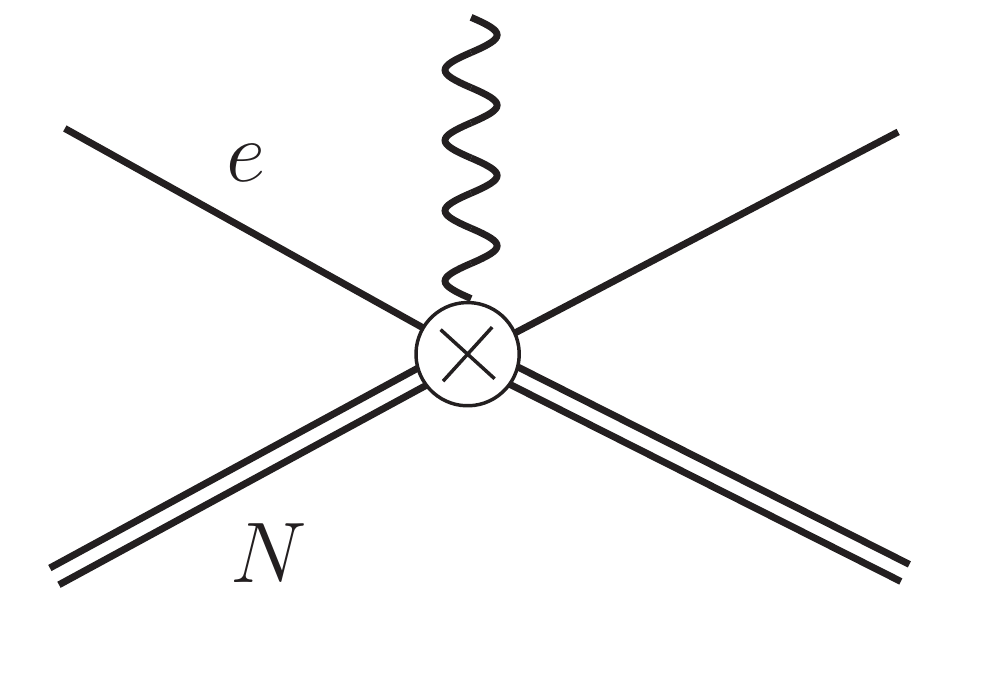}
\includegraphics[viewport=160 547 450 720, clip=true, scale=0.4]{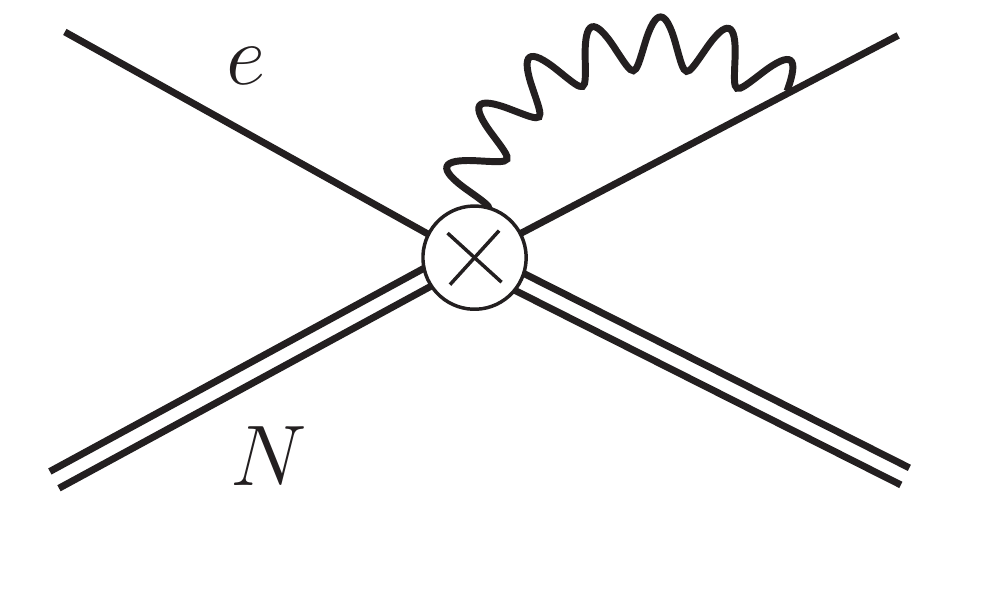}
\caption{Left panel: schematic representation of the contact electron-nucleon-photon operator ${\cal O}_9$. 
Right panel: integrating out the hard photon loop transmutes this operator to the more familiar four-fermion 
operator $\bar e i \gamma_5 e \bar NN$. } 
\label{fig4}
\end{figure}

The coefficient $C_9$ already incorporates the $CP$-odd photon exchange associated with the penguin vertex, and as a second step
one can `integrate out' the remaining photon (see Fig.~\ref{fig4}). At 1-loop order, the 
$\tilde F_{\mu\nu} (\bar e \gamma_\nu e)$ part of $O_9$ transmutes to $\bar e \gamma_\mu \gamma_5e$,
with a quadratically divergent loop integral. The $(\bar e \gamma_\mu \gamma_5e)\partial_\mu (\bar NN)$ operator
can then be reduced to the standard $C_{SP}$ form upon using the divergence of the axial current. Performing this 
computation, we can identify 
\be
\label{matching}
\frac{G_F}{\sqrt{2}} \times C_{SP} \simeq    \frac{3 \alpha m_e}{2\pi}  \int dk^2 C_9.
\ee
In this expression, the quadratic divergence will be cut off by the hadronic vertex 
form factor $C_9(k^2)$. In general, $C_9$ will depend independently on both $k_0^2$ and $\vec{k}^2$. In fact, in deriving (\ref{matching}) we have used 
$k_\alpha k_\beta d^4k \to \fr14 g_{\alpha\beta} k^2 d^4k$, which will not strictly be valid in the presence of a non-Lorentz invariant form factor. Nonetheless, in relying on the hadronic form factor to cut off the divergent integral, we are necessarily focusing only on an order of magnitude estimate, and thus we can safely neglect these issues.

\subsection{EM penguin contrubution to \boldmath{$C_9$}}

We now estimate the CKM-induced value of $C_9$ and the resulting value of 
$C_{SP}$ by combining the $\Delta S =1$ and $\Delta S = -1$ transitions. Inside the 2PE box diagram (Fig.~3a), 
there are a multitude of propagating hadronic states with non-zero strangeness. To simplify the computation, 
we shall saturate them with the $\Sigma$-baryon, and estimate the resulting $\Sigma N \gamma$ vertices 
using the {\em on-shell} data for $\Sigma$ decays. Once again, this approximation will only capture part of the answer, as the off-shell
vertex may contain further contributions not captured by $\Sigma$ decays. However, this simplification will be sufficient for the purpose of 
an order-of-magnitude estimate.

We start by quoting the result for the $CP$-violating part of the 
electromagnetic penguin operator,
\begin{equation}
{\cal L}_{\rm pen} = iC_{\rm pen} (\bar s_L \gamma_\mu d_L)(\bar e \gamma_\mu e)+(h.c.),
\end{equation}
where  \cite{em_pen}
\begin{equation}
C_{\rm pen}=  \frac{G_F}{\sqrt{2}}\times (s_1s_2s_3c_2)\sin\delta\times\frac{ 4 \alpha}{9 \pi} \times 
\log\left(\frac{m_W^2}{m_c^2}\right).
\end{equation}
The logarithmic factor originates from the relative sign between the charm and top contributions in the loop, 
and with logarithmic accuracy, the upper limit can be chosen to be $m_W$. The $i$ in front of the whole expression is the 
signature of  $CP$-violation. 

In order to go from $\bar s_L \gamma_\mu d_L$ to the sigma-nucleon vertex, one can 
use experimental information from semi-leptonic nucleon and hyperon decays and $SU(3)$ flavor 
symmetry (see {\em e.g.} \cite{German}),
\be
\langle \Sigma | \bar s_L \gamma_\mu d_L|p\rangle \simeq 
- 0.8 \bar\Sigma( \gamma_\mu - 0.43\gamma_\mu \gamma_5)p,
\ee
with analogous relations for the neutron. However, retaining the exact isospin factors and the relation
$g_A/g_V = 1.26$ is beyond the precision goal of the present estimate. Note that at other stages of the calculation ({\em e.g.}   
saturating the integral $\int C_9 dk^2$ by its upper limit) the precision is not better than 
an order-of-magnitude. Therefore, we will simply assume that $\bar s_L \gamma_\mu d_L$ generates the $\sim \bar \Sigma_L \gamma_\mu N_L$
Lorentz structure with an order unity coefficient, and disregard the isospin dependence and any moderate deviations of the current matrix elements 
from unity. Thus, to obtain an estimate, we simply assume the approximate transition 
\be
{\cal L}_{\rm pen} \to iC_{\rm pen} (\bar \Sigma_L \gamma_\mu N_L)(\bar e \gamma_\mu e)+(h.c.).
\ee

The tree level amplitude for the $\Sigma N\gamma$ transition can be extracted directly from 
the $\Sigma \to p\gamma$ decay, modulo the off-shell virtuality of both $\Sigma$ and $\gamma$. 
Following Ref.~\cite{German}, we write
\be
{\cal L}_{\rm tree} \simeq \frac{eG_F}{2}\bar \Sigma \sigma_{\mu\nu} (a +b\gamma_5)pF_{\mu\nu},
\ee
where $a$ and $b$ are phenomenological functions of $k^2$, the momentum of the remaining photon, which have both real and imaginary (dispersive) parts. We will only keep track of  the substantially larger real parts, which can be fixed by the total decay rate and angular correlations in the decay. Note that, as defined, 
$a$ and $b$ incorporate some dependence on the CKM angles, so it is useful to introduce CKM angle-free functions $A$ and $B$ via
$a = s_1c_1c_3 \times A$ and $b = s_1c_1c_3 \times B$.
The coefficient of the left-handed structure is given by the combination $(a-b)$, 
which is determined by the relations \cite{German},
\begin{align}
&a(0)^2+b(0)^2 \simeq (15~{\rm MeV})^2, \nonumber\\
 & a(0)b(0) \simeq -85~{\rm MeV}^2.
\end{align}
This data implies $a(0)-b(0) \simeq 20$ MeV, and thus $A(0)-B(0)) \simeq 100$ MeV, which is a natural energy scale in this problem. 

We can now combine the $CP$-odd and $CP$-even vertices in ${\cal L_{\rm pen}}$ and ${\cal L_{\rm tree}}$ respectively in one diagram containing an 
intermediate $\Sigma$ state (see Fig.~3a). Expanding the internal $\Sigma$ propagator to first order in the small momentum transfer, we can isolate the Lorentz structure corresponding to  the nucleon-spin-independent term. Carrying out this procedure, we observe that the $O_9$ operator is indeed generated, and the matching procedure gives the following Wilson coefficient in the limit of small photon momenta,
\begin{eqnarray}
C_9(0) \simeq C_{\rm pen} \times \frac{G_F(a(0)-b(0))}{2(m_\Sigma^2-m_N^2)}.
\end{eqnarray}

Ideally, one would need the full dependence of $C_9$ on photon virtualities in order to compute the loop integral in (\ref{matching}). For the 
purpose of obtaining an estimate, we take
\be
\int C_9 dk^2 \sim C_{\rm pen} \times \frac14G_F(a(0)-b(0)),
\ee
which corresponds to setting the cutoff of the $dk^2$ integral at $\fr12(m_\Sigma^2-m_N^2) \sim (500~{\rm MeV})^2$.
This choice seems justified, since it is of order the characteristic quark momenta inside nucleons. 

We are now ready to combine all the numerical factors in a final estimate of $C_{SP}$,
\begin{align}
C_{SP} &\sim {\cal J} \times \frac{\alpha^2}{6 \pi^2} \times  {G_F m_e(A(0)-B(0))}\times \log\left(\frac{m_W^2}{m_c^2}\right)\nonumber\\
 &\sim 10^{-19},
 \end{align}
leading to the equivalent electron EDM benchmark,
\begin{equation}
d_e^{\rm equiv}({\cal J}) = r C_{SP}({\cal J}) \sim 10^{-39} e{\rm cm},
\end{equation}
which is roughly one order of magnitude below our limiting estimate (\ref{est}) if one uses 
$m_{\rm had}\sim 100$ MeV, but many orders of magnitude above 
$d_e({\cal J})$ proper. Given the approximate nature of this estimate, it is certainly possible that the full equivalent electron EDM may reach the level of the scaling estimate, 
$10^{-38}e$cm.

\section{Concluding Remarks}

The CKM benchmark obtained above is $\sim 9-10$ orders of magnitude below the best current sensitivity to the electron EDM 
from the ThO experiment. This gap is not a surprise, given the high degree of suppression for all CKM-induced contributions 
to flavor-conserving observables. Nonetheless, it is somewhat larger than in other channels, such as the neutron or diatomic EDMs, for which the CKM contributions are $\sim 5-6$ orders of magnitude below current sensitivity. Since current electron EDM measurements 
are performed with atoms and molecules, rather than isolated electrons, the $CP$-odd observable 
measured in these experiments is inevitably sensitive not only to $d_e$ itself, but also to nucleon-spin-independent four-fermion operators, characterized by $C_{SP}$. We have  estimated the value of $C_{SP}$ induced by the CKM phase, and inferred the size of the {\em equivalent}
electron  EDM, $d_e^{\rm equiv} \equiv rC_{SP}({\cal J}) \sim 10^{-38}e$cm.  This is many orders of magnitude {\em larger} 
than the contribution of $d_e({\cal J})$ proper, which itself  is highly 
suppressed by the degree of flavor cancellations within the 
closed quark loops of Fig.~1. 

The dominance of the CKM contribution to $C_{SP}$ over $d_e$, by $\sim 5$ orders of magnitude, shows 
how important the relative contribution of $C_{SP}$ can be. Other known examples  where $C_{SP}$ may also provide a dominant contribution include some 
special cases of beyond the SM physics, such as the two-Higgs doublet model and supersymmetric models at large $\tan\beta$ \cite{Barr,LebedevPospelov,Demir03}. 

Finally, given that the field content of the SM needs to be enlarged to include the effects of non-zero neutrino masses, 
one can ask about the size of $d_{e}$ and $C_{SP}$ in the SM extended by right-handed neutrinos. This is, in a certain sense, the most conservative extension of the SM that is supported by experimental evidence. The Majorana nature of neutrino masses then allows for a non-zero $d_e$ at the two-loop level
\cite{NgNg,Archambault,deGouvea}. However, the typical size of this contribution remains extremely small, 
as it is suppressed by the smallness of the neutrino Yukawa couplings and/or by the large scale for the 
Majorana mass of the right-handed neutrinos. In exceptional/tuned cases $d_e$ can reach $10^{-33}e$cm
\cite{Archambault}, which is still below
the current sensitivity limit \cite{newlimit}.

\begin{acknowledgments}
The work of  M.P. and A.R. is supported 
in part by NSERC, Canada, and research at the Perimeter Institute is supported in part by the Government 
of Canada through NSERC and by the Province of Ontario through MEDT. 
\end{acknowledgments}

\end{document}